\documentclass[aps,letterpaper,twocolumn,prl,showpacs]{revtex4-1}
\usepackage{amsmath,bm}
\usepackage{graphicx,color}
\usepackage[english]{babel}
\usepackage{hyperref}

\begin{document}

\title{Origins of clustered frequency combs in Kerr microresonators}

\author{Noel~Lito~B.~Sayson$^{1,2}$}
\author{Hoan~Pham$^{1}$}
\author{Karen~E.~Webb$^{1}$}
\author{Vincent Ng$^{1}$}
\author{Luke~S.~Trainor$^{3}$}
\author{Harald~G.~L.~Schwefel$^{3}$}
\author{St\'{e}phane~Coen$^{1}$}
\author{Miro~Erkintalo$^{1}$}
\author{Stuart~G.~Murdoch$^{1}$}
\email{s.murdoch@auckland.ac.nz}

\affiliation{$^{1}$The Dodd-Walls Centre for Photonic and Quantum Technologies, Department of Physics, The University of Auckland, Auckland 1142, New Zealand \\
$^{2}$Physics Department, Mindanao State University - Iligan Institute of Technology, Tibanga, 9200 Iligan City, Philippines \\
$^{3}$The Dodd-Walls Centre for Photonic and Quantum Technologies, Department of Physics, University of Otago, 730 Cumberland Street, Dunedin 9016, \mbox{New Zealand}
}

\begin{abstract}
Recent experiments have demonstrated the generation of widely-spaced parametric sidebands that can evolve into ``clustered'' optical frequency combs in Kerr microresonators. Here we describe the physics that underpins the formation of such clustered comb states. In particular, we show that the phase-matching required for the initial sideband generation is such that (at least) one of the sidebands experiences anomalous dispersion, enabling that sideband to drive frequency comb formation via degenerate and non-degenerate four-wave mixing. We validate our proposal through a combination of experimental observations made in a magnesium-fluoride microresonator and corresponding numerical simulations. We also investigate the coherence properties of the resulting clustered frequency combs. Our findings provide valuable insights on the generation and dynamics of widely-spaced parametric sidebands and clustered frequency combs in Kerr microresonators.
\end{abstract}

\maketitle

\noindent Thanks to their remarkably high finesse and small modal area, optical microresonators allow for the efficient generation of new frequencies through nonlinear interactions~\cite{harald1}. In particular, they enable the creation of unique optical frequency combs \cite{delhaye07,herr14,yi15}, whose potential applications range from waveform synthesis \cite{ferdous11} to telecommunications \cite{pfeifle14}. These frequency combs arise via Kerr nonlinear interactions when the resonator is driven in the regime of anomalous group-velocity dispersion (GVD): phase-matched four-wave mixing (FWM) drives a modulation instability (MI) that results in the growth of symmetrically-detuned sidebands.

Besides optical frequency combs, FWM in Kerr microresonators has more recently been shown to permit the generation of isolated pairs of large frequency shift parametric sidebands~\cite{matsko16,fujii17,liang15,huang17,sayson17}. Such sidebands typically arise when the resonator is driven in the regime of normal GVD, with phase-matching enabled by interactions between different mode families~\cite{liang15} or higher-order dispersion~\cite{matsko16,sayson17,fujii17}. In particular, as is well-known in the context of nonlinear fiber optics, the presence of a negative fourth-order dispersion can compensate for positive (normal) GVD. In stark contrast to the standard MI in the anomalous dispersion regime, pumping in the normal dispersion regime typically results in the generation of sidebands that exhibit very large frequency shifts and that can be widely tuned by small changes in the pump wavelength. In the context of microresonators, fixed sideband shifts of $\sim 40$~THz have been observed in both MgF$_2$ \cite{matsko16} and silica resonators \cite{fujii17}, while wideband tunability across 700~nm has been demonstrated in a silica microsphere \cite{sayson17}.

Because the generation of large-frequency shift parametric sidebands is underpinned by a single phase-matched FWM process, one may expect the sidebands to retain the high spectral purity of the pump. However, experimental studies have shown that, for certain parameter configurations, ``clustered'' frequency combs may form around the pump and the primary sidebands~\cite{matsko16,fujii17}. Whilst the ability to generate localized frequency comb structures far from the pump frequency can prove useful for some applications, for others --- such as standard spectroscopic measurements requiring a single optical frequency --- the process may be unwanted. Either way, there is interest to gain a detailed understanding of the physics and dynamics that underlie the formation of clustered frequency combs.

\begin{figure*}[htb]
\centering
\includegraphics[width=\linewidth]{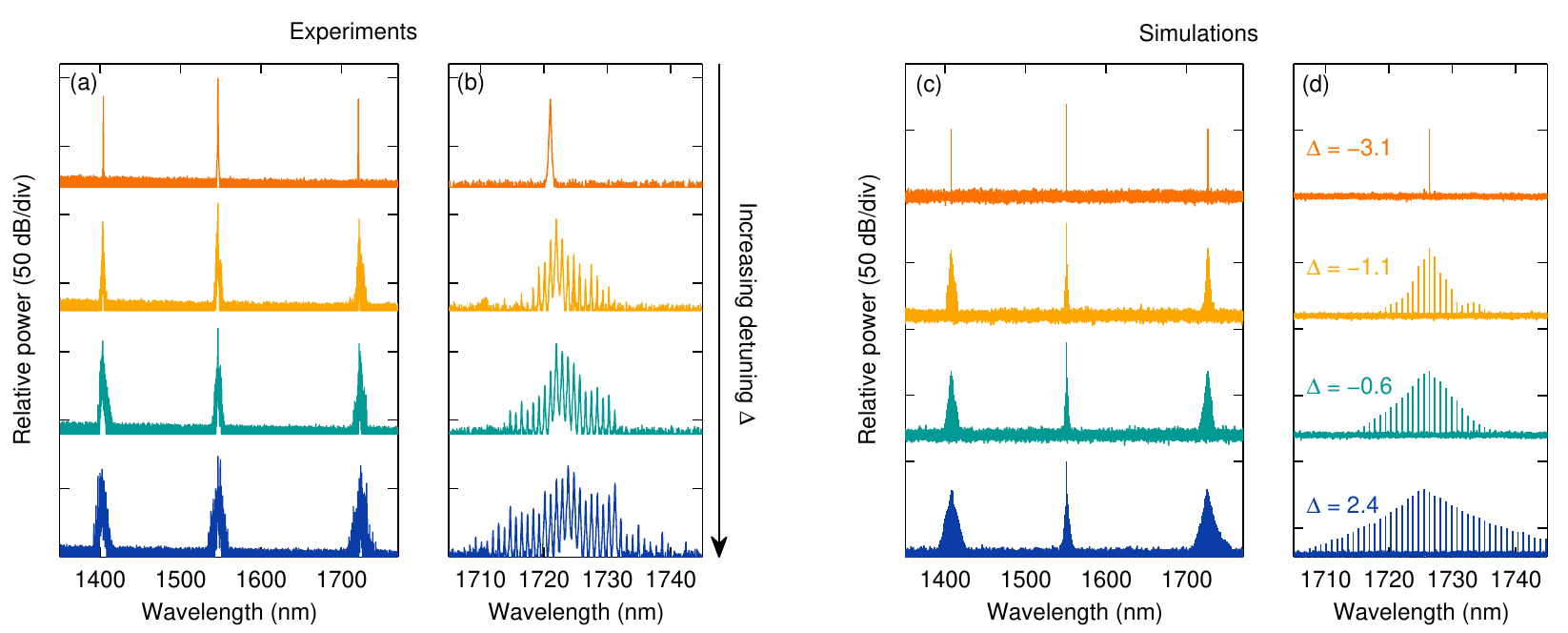}
\caption{\small (a) Experimentally measured spectra for increasing pump detuning $\Delta$, showing the formation of clustered frequency combs. (b) Zoom of the long-wavelength sideband for the spectra shown in (a). (c), (d) Corresponding results from numerical simulations of the generalized LLE. To mimic the optical spectrum analyzer used to acquire the experimental data shown in (a), (b), the simulated spectra were averaged over several photon lifetimes and a noise floor at $-70$~dB was added to the simulated data. Simulations use detunings $\Delta$ as indicated, with other parameters as follows: $\beta_{2,\mathrm{p}} = 0.486~\mathrm{ps^2/km}$; $\beta_{3,\mathrm{p}} = 0.07~\mathrm{ps^3/km}$; \mbox{$\beta_{4,\mathrm{p}} = -4\times10^{-4}~\mathrm{ps^4/km}$}; $\gamma = 1~\mathrm{W/km}$; $L = 2.5~\mathrm{mm}$; $P_\mathrm{in} = |E_\mathrm{in}|^2 = 90~\mathrm{mW}$; $\alpha = \theta = 7.3\times 10^{-5}$.}
\label{fig1}
\vskip-5pt
\end{figure*}

In this Letter, we present a simple explanation for the appearance of clustered frequency combs in Kerr microresonators. Specifically, we show that the phase-matching that underpins the generation of large-frequency shift parametric sidebands is such that (at least) one of the sidebands is always generated in the regime of anomalous GVD, enabling it to act as a pump in standard Kerr comb generating processes. We also investigate the coherence characteristics of the resulting frequency comb clusters. By elucidating the origins and dynamics of clustered frequency combs, our findings could facilitate the development of applications where such clusters may play a role.

We begin by presenting experimental observations of cluster formation in an optical microresonator. Our experiments use a crystalline MgF$_2$ microresonator, fabricated by diamond point turning and polished to a high optical quality with diamond abrasives similar to ~\cite{harald2,harald3}. The resonator has a diameter of $800~\mu$m (free spectral range $\textrm{FSR}=87.1$~GHz) and a measured finesse \mbox{$\mathcal{F}=4.3\cdot10^4$}. We drive the resonator with a tunable C-band external cavity laser at 1547~nm, amplified with an erbium-doped fiber amplifier and filtered with a 1~nm bandpass filter to produce a 90~mW continuous wave pump. The pump is coupled to the microresonator via a tapered silica optical fiber with a waist diameter of approximately $1~\mu$m. The taper output is split three ways: 50\% of the output is monitored on an optical spectrum analyzer, 5\% is sent to an electronic spectrum analyzer for RF noise measurements, and the remaining signal is passed through an offset bandpass filter (pass band 1200--1400~nm) and detected on an amplified photodiode. This filtered channel allows us to isolate nonlinear signals originating from large frequency shift sidebands, and thus to identify resonances where such sidebands are generated. We then finely tune the pump wavelength into the desired cavity resonance by piezo tuning while continuously monitoring the optical spectrum.

Figure~\ref{fig1}(a) shows selected experimentally measured spectra at the microresonator output for increasing values of normalised detuning $\Delta$, with \mbox{$\Delta \approx \mathcal{F}\cdot(\omega_k-\omega_p)/(2\pi\text{FSR})$}, where $\omega_k$ is the angular frequency of the cavity mode closest to the pump frequency $\omega_\mathrm{p}$. At the onset of parametric oscillation, a single pair of widely-detuned sidebands is observed, with a frequency shift of approximately 19.7~THz. As the pump is scanned further into the resonance, small frequency comb clusters with a single FSR spacing are generated around the pump and the two primary sidebands. The size of these clusters increases as the pump detuning is further increased, as highlighted in Fig.~\ref{fig1}(b), which shows a zoom of the long-wavelength clusters.

Our experimental observations are in good agreement with corresponding numerical simulations based on the generalized Lugiato-Lefever equation (LLE)~\cite{coen13a}:
\begin{equation}
\begin{split}
t_\textrm{R}\frac{\partial E(t,\tau)}{\partial t}=&\left[-\alpha(1 + i\Delta)+iL\sum_{k\geq2}\frac{\beta_{k,\mathrm{p}}}{k!}\left(i\frac{\partial}{\partial\tau}\right)^k\right]E\\
&+i\gamma L|E|^2E+\sqrt{\theta}E_\textrm{in}.
\end{split}
\label{lle}
\end{equation}
Here, $E(t,\tau)$ is the intracavity field envelope, $t$ and $\tau$ are the slow and fast time variables, respectively, $t_\textrm{R}=1/\textrm{FSR}$ is the roundtrip time, $\alpha=\pi/\mathcal{F}$ is half the total power loss per roundrip, $\beta_{k,\mathrm{p}}$ are the dispersion coefficients evaluated at the pump frequency $\omega_\mathrm{p}$, $\gamma$ is the nonlinear coefficient, $L$ is the round trip length of the resonator, and $\theta$ is the power coupling coefficient. Figures~\ref{fig1}(c) and (d) show numerically simulated spectra at selected cavity detunings with parameters similar to our experiments (see caption). As can be seen, our simulations reproduce the experimentally observed behaviour: a single pair of widely-detuning sidebands appear for low detunings, then as the detuning increases, clustered frequency combs form around the pump and the primary sidebands.

To explain the observed cluster formation, we first recall the phase-matching condition that underpins the primary widely-detuned parametric sidebands. Truncating the dispersion at fourth order (i.e., $\beta_{k,\mathrm{p}} = 0$ for $k>4$), phase-matching is achieved at sideband frequencies $\omega_\pm$ satisfying~\cite{sayson17}
\begin{equation}
\frac{\beta_{4,\mathrm{p}}}{24}\Omega^4 + \frac{\beta_{2,\mathrm{p}}}{2}\Omega^2 + 2\gamma P_\mathrm{p} - \alpha\frac{\Delta}{L} = 0,
\label{PM}
\end{equation}
where $\Omega$ describes the sidebands' frequency shift from the pump ($\omega_\pm = \omega_\mathrm{p}\pm\Omega$) and $P_\mathrm{p}$ is the intracavity power at the pump wavelength. To first order, the last two terms on the right-hand of Eq.~\eqref{PM} can be neglected, yielding:  $\Omega_\textrm{pm}^2 = -12\beta_{2,\mathrm{p}}/\beta_{4,\mathrm{p}}$. Considering now the local GVD experienced by the two phase-matched sidebands, we find:
\begin{equation}
\beta_{2}(\omega_\pm) = -5\beta_{2,\textrm{p}} \pm \beta_{3,\textrm{p}}\sqrt{-\frac{12\beta_{2,\textrm{p}}}{\beta_{4,\textrm{p}}}}.
\label{beta2}
\end{equation}

Large frequency shift sidebands occur when $\beta_{2,\textrm{p}}$ is positive (normal dispersion) and $\beta_{4,\textrm{p}}$ is negative. As a consequence, a simple analysis of Eq.~\eqref{beta2} shows that the local $\beta_2$ experienced by at least one of the two phase-matched sideband frequencies must be negative (i.e., anomalous). This holds true irrespective of the sign or magnitude of $\beta_{3,\textrm{p}}$, permitting a simple interpretation for the formation of clustered frequency combs. Specifically, the sideband that experiences anomalous GVD may act as a pump for a secondary phase-matched MI process, amplifying its own small-frequency shift MI sidebands. The peak gain per roundtrip of the sideband field is $g = \gamma P L$, where $P$ is the intracavity power of the primary sideband experiencing anomalous dispersion; if this gain exceeds the per roundtrip loss ($g>\alpha$), standard anomalous-dispersion comb formation dynamics will take place. Once sidebands have formed around the primary sideband experiencing anomalous dispersion, further FWM will transfer the resulting combs to the other sideband and the pump.

To test our theory, we repeated our simulations for a wide range of detunings, examining whether the onset of cluster formation is correlated with the point at which the power of the anomalous-dispersion sideband exceeds the MI threshold \mbox{$P_\mathrm{MI} = \alpha/(\gamma L)$}.  For the particular dispersion coefficients used, we find that it is the low-frequency (red-detuned) sideband that experiences anomalous dispersion ($\beta_2(\omega_-)=-10.5$~ps$^2$km$^{-1}$). We also note that, as the gain bandwidth of the initial FWM process (responsible for the large frequency shift primary sidebands) can encompass several cavity modes, the field in the anomalous dispersion regime can be temporally modulated; MI is to be expected once the \emph{peak} power of that field exceeds threshold.

\begin{figure}[!t]
\centering
\includegraphics[width=\linewidth]{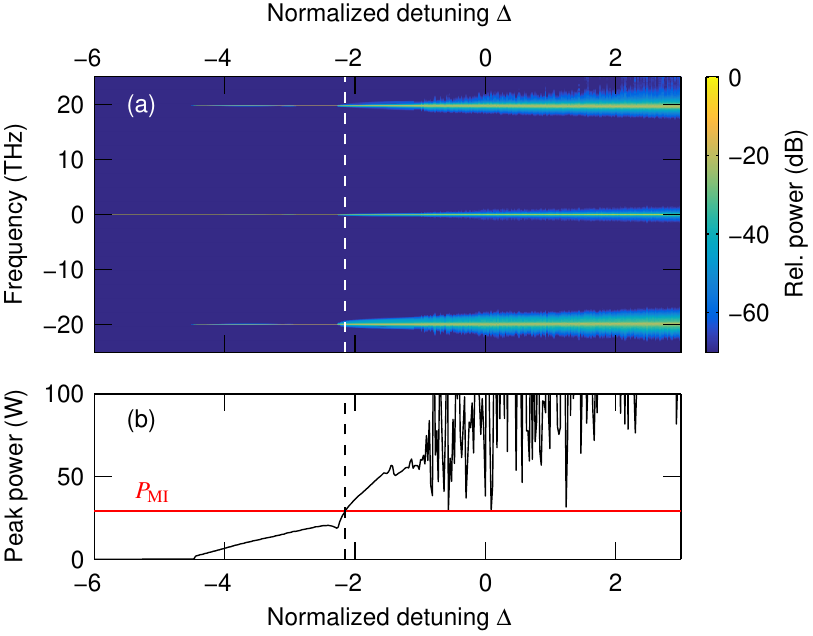}
\caption{\small (a) Numerically simulated spectra at the microresonator output as a function of cavity detuning $\Delta$. (b) Peak intracavity power of the red-detuned sideband (black solid line) and the MI threshold (red solid line). Dashed vertical lines in (a) and (b) indicate the detuning at which the sideband peak power is equal to the MI threshold.}
\label{fig2}
\vskip-10pt
\end{figure}

In Fig.~\ref{fig2}(a), we show numerically simulated cluster spectra as a function of cavity detuning, whilst Fig.~\ref{fig2}(b) shows the corresponding peak intracavity power of the field in the anomalous dispersion regime, obtained by spectrally isolating the components around the low-frequency sideband with a numerical spectral filter. Also shown in Fig.~\ref{fig2}(b) is the MI threshold power $P_\mathrm{MI}$. We see that, at a detuning of approximately $\Delta\approx-2.1$, the peak power of the anomalous sideband first crosses the MI threshold. An examination of Fig.~\ref{fig2}(a) shows that this detuning agrees very well with the point at which cluster formation first occurs, thus corroborating our simple theory.

\begin{figure}[t]
\centering
\includegraphics[width=\linewidth]{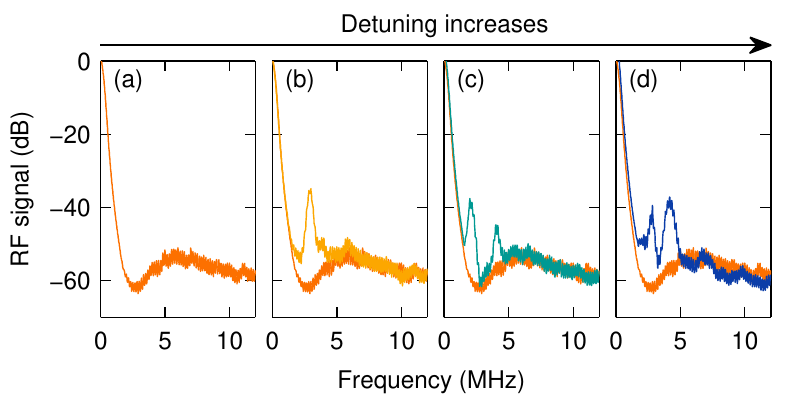}
\caption{\small Experimentally measured low-frequency RF noise at four different detunings, with the detuning increasing from (a) to (d) as indicated. The corresponding optical spectra are shown in Fig.~\ref{fig1}(a). The RF spectrum measured for the lowest detuning (orange curve) is shown in all panels to facilitate comparison.}
\label{RF}
\end{figure}

We have also investigated the stability of the generated frequency comb clusters. In Fig.~\ref{RF}, we show results from experimental RF intensity noise measurements corresponding to the different spectra shown in Fig.~\ref{fig1}(a). As can be seen, when only a single parametric sideband is present [Fig.~\ref{RF}(a)], we observe no excess noise, indicating that these sidebands are highly coherent. In contrast, the generation of frequency comb clusters is found to be accompanied by a clear increase in the measured low-frequency intensity noise [Fig.~\ref{RF}(b)--(d)] --- a known signature of unstable comb dynamics~\cite{papp11, herr12}. To gain more insights, we have examined the coherence of numerically simulated comb clusters for a range of detunings [other parameters as in Fig.~\ref{fig1}(b)]. Following the approach explained in~\cite{erkintalo14,torres, webb16}, we evaluate the complex degree of first order coherence $g_{12}^{(1)}(\Delta t)$ for each of the spectral components over a wide range of coherence delays $\Delta t$. In what follows, we focus our analysis in the spectral region around the anomalous-dispersion (red-detuned) parametric sideband, yet note that qualitatively similar behaviour is observed around the pump and the other sideband.

\begin{figure}[!t]
\centering
\includegraphics[width=\linewidth]{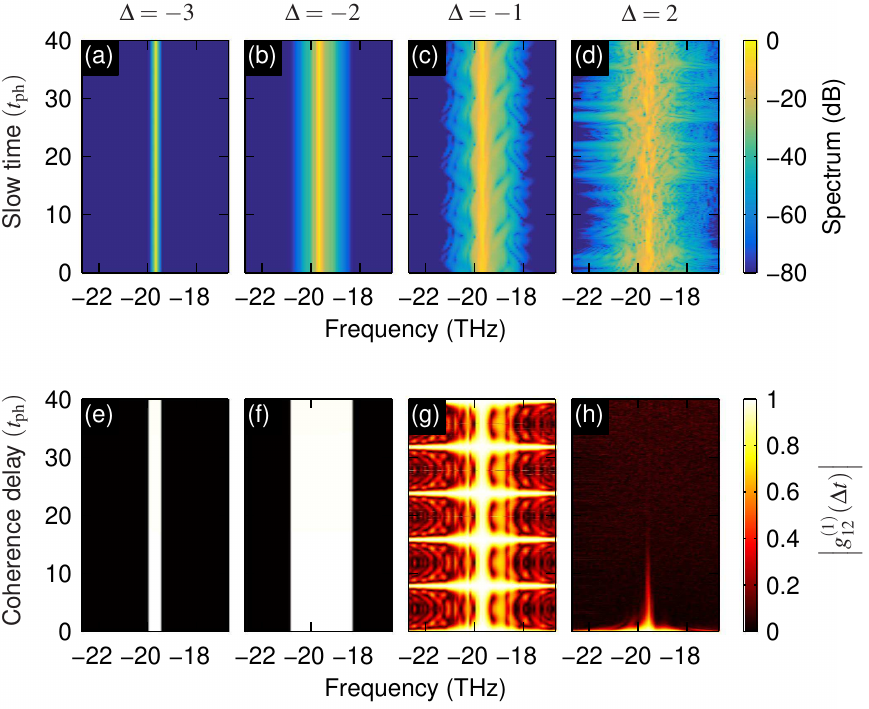}
\caption{\small (a)--(d) Numerically simulated evolution of the frequency comb cluster around the anomalous dispersion sideband. (e)--(h) Corresponding degree of coherence as a function of the coherence delay. The simulations ware carried out at normalized detunings of (a), (e) $\Delta=-3$, (b), (f) $\Delta=-2$, (c), (g) $\Delta=-1$, and (d), (h) $\Delta=2$. Other parameters as in Fig.~\ref{fig1}.}
\label{fig3}
\vskip-10pt
\end{figure}

In our simulations, we have observed four different regimes of cluster coherence, and in Fig.~\ref{fig3}, we show illustrative examples of the different behaviours. Specifically, Figs.~\ref{fig3}(a)--(d) show spectral evolutions of numerically simulated comb clusters over 40 photon lifetimes at normalized detunings of $\Delta=-3$, $\Delta=-2$, $\Delta=-1$, and $\Delta=2$, respectively, while Fig.~\ref{fig3}(e)--(h) shows the corresponding degrees of coherence. At a detuning of $\Delta=-3$, only a single frequency component --- corresponding to the parametric sidebands generated by the pump --- is present [Fig.~\ref{fig3}(a)]. This line exhibits high coherence [Fig.~\ref{fig3}(e)]. At a slightly larger detuning of $\Delta=-2$, we observe clusters [Fig.~\ref{fig3}(b)] that remain unchanged from roundtrip-to-roundtrip; all of the comb lines exhibit perfect coherence. At a larger detuning of $\Delta = -1$, we find that the cluster exhibits periodic breathing behaviour, with a period of about 8 photon lifetimes. Similarly to breathing cavity solitons~\cite{erkintalo14}, the degree of coherence of such a cluster is periodically restored for increasing delays, with the period of recurrence coinciding with the breathing period. Finally, for a larger detuning of $\Delta=2$, we find clusters that exhibit chaotic fluctuations [Fig.~\ref{fig3}(d)]. In this regime, the degree of coherence rapidly degrades over most of the spectrum as $\Delta t$ increases [Fig.~\ref{fig3}(h)]. Although a detailed study is beyond the scope of our present work, our simulations suggest that this fully incoherent regime is the most prevalent mode of cluster operation, manifesting itself over the widest range of cavity detunings. This finding could explain why all the clusters we have observed in our experiments display elevated low-frequency RF noise.

As might be expected, the coherence dynamics reported above are quite similar to those observed in a standard microresonator comb driven by an external pump in the anomalous dispersion regime~\cite{herr12}. The primary point of difference is that, in our case the anomalous-dispersion pump is being generated internally via another FWM process. It is also worth noting that, for different dispersion parameters, we have observed the formation of coherent clustered combs where the cluster spacing is more than one FSR. Again, this behaviour is very reminiscent of standard comb formation dynamics in the anomalous dispersion regime.

Before closing, we briefly discuss possible alternative routes for the generation of clustered frequency combs as well as their dynamics. Specifically, additional simulations (not shown here) reveal that, for some parameter configurations, narrow clusters can arise even before the peak power of the anomalous dispersion sideband exceeds the MI threshold. This can be explained by the fact that, as noted above, the initial FWM process can drive the amplification of several adjacent cavity modes: non-degenerate FWM between the amplified spectral components can result in the generation of new components spaced by a single FSR. Similarly to comb generation via bichromatic pumping~\cite{hansson14}, this mechanism is not subject to any power thresholds; however, we find that clusters generated in this way, at powers below the MI threshold, are small and contain very few lines with appreciable power. Our simulations also show that, irrespective of their generating mechanism, the cluster bandwidth does not always increase monotonically with detuning. This can be understood in light of the detuning-dependence of the precise phase-matching that underpins the initial FWM process. Specifically, a change in detuning can shift the peak of the FWM gain away from a cavity resonance, reducing the conversion efficiency of the initial FWM process, and hence the power of the sideband that acts as a pump in the cluster generating process. A detailed study of such dynamics is left for future work.

In conclusion, we have studied the formation of clustered frequency combs in a Kerr nonlinear optical microresonator. We have shown that the phase-matching that underpins the generation of large-frequency shift parametric sidebands is such that at least one of the generated sidebands experiences anomalous dispersion, enabling it to act as a pump in subsequent Kerr comb generation processes. We have also investigated the coherence of the frequency comb clusters, finding experimental and numerical evidence of not only coherence degradation, but also more complex oscillatory behaviours. In addition to providing new insights into the origin of clustered frequency combs, we expect that the ability to predict whether cluster formation will occur for particular parameters will prove useful for the future design of widely tunable Kerr parametric oscillators.

We acknowledge support from the Marsden Fund and Rutherford Discovery Fellowships of the Royal Society of New Zealand.

\newcommand{\enquote}[1]{``#1''}

\end{document}